\documentstyle[multicol,prd,aps,epsf]{revtex}

\newcommand{\ve}[1]{ {\bf #1}}
\newcommand{\omg}{(m\Omega)}
\newcommand{\Eadm}{E_{\mbox{\rm\tiny ADM}}}
\newcommand{\pps}{\mbox{\small$++$}}
\newcommand{\pms}{\mbox{\small$+-$}}
\newcommand{\mms}{\mbox{\small$--$}}

\begin{document}
\draft

\title{Quasi-circular Orbits for Spinning Binary Black Holes}

\author{Harald P. Pfeiffer and Saul A. Teukolsky}
\address{Department of Physics, Cornell University, Ithaca, New York\ \ 14853}
\author{Gregory B. Cook}
\address{Department of Physics, Wake Forest University, Winston-Salem,
         North Carolina, 27109}

\date{\today}

\maketitle

\begin{abstract}
Using an effective potential method we examine binary black holes
where the individual holes carry spin.  We trace out sequences of
quasi-circular orbits and locate the innermost stable circular orbit
as a function of spin. At large
separations, the sequences of quasi-circular orbits match well with
post-Newtonian expansions, although a clear signature of the
simplifying assumption of conformal flatness is seen. The position of
the ISCO is found to be strongly dependent on the magnitude of the
spin on each black hole.  At close separations of the holes, the
effective potential method breaks down.  In all cases where an ISCO
could be determined, we found that an apparent horizon
encompassing both holes forms for separations well inside the ISCO.
Nevertheless, we argue that
the formation of a common horizon is still associated with the
breakdown of the effective potential method.
\end{abstract}


\begin{multicols}{2}
\section{Introduction}

The inspiral and coalescence of binary black hole systems is a prime
target for upcoming gravitational wave detectors such as LIGO.  Such
systems will be circularized by the emission of gravitational waves,
and will evolve through a quasi-equilibrium sequence of circular
orbits. At the innermost stable circular orbit (ISCO) we expect a
transition to a dynamically plunging orbit. It is anticipated that this
transition will impart a characteristic signature on the gravitational
waveform. It is therefore important to know the orbital frequency at
the ISCO, since the corresponding gravitational wave frequency is
predominantly just twice this frequency.

Predicting the waveform in detail from the transition at the ISCO to
the final merger requires the full machinery of numerical
relativity. These calculations require appropriate initial data. Out of
the large space of solutions of the initial-value equations of general
relativity, we need an algorithm to select solutions corresponding to
black holes in quasi-circular orbits. The effective potential method
\cite{Cook:1994:TID} allows one to construct such solutions, and to
determine the properties of the ISCO.

The effective potential is based on the fact that minimizing the
energy of a system yields an equilibrium solution. This follows from
the Hamiltonian equations of motion: If the Hamiltonian $\cal H$ is
minimized with respect to a coordinate $q$ and a momentum $p$, then
$\dot{q}=\partial{\cal H}/\partial p=0$ and $\dot{p} =
-\partial {\cal H}/\partial q=0$.  The energy of two objects in
orbit about each other can be lowered by placing the objects at rest
at their center of mass. Therefore minimizing the energy with respect
to all coordinates and momenta will not yield a circular orbit. To
find circular orbits in Newtonian gravity, one can minimize the energy
while holding the angular momentum constant. This procedure works as
well for a test-mass orbiting a Schwarzschild black hole, where one
minimizes the ADM energy. This can be seen as follows. For geodesic
motion, one finds \cite{Wald:1984:GR}
\begin{equation}\label{eqn:Schwarzschild}
\frac12\dot r^2 + \frac12\left(1-\frac{2M}{r}\right)
	\left(\frac{\tilde{L}^2}{r^2}+1\right) = \frac12\tilde{E}^2.
\end{equation}
Here $M$ is the mass of the black hole, $\tilde{E}$ is the energy per
unit rest mass of the test-particle as seen from infinity and
$\tilde{L}$ its orbital angular momentum per unit rest mass. Denote
the rest-mass of the test-particle by $M'$. Then the ADM energy is simply
$\Eadm=M+\tilde{E}M'$, and minimizing $\Eadm$ is equivalent to
minimizing $\tilde{E}$. Hence minimizing the left hand side of
(\ref{eqn:Schwarzschild}) with respect to $r$ yields the radius of
circular orbits as a function of angular momentum. Minimization of
(\ref{eqn:Schwarzschild}) with respect to $\dot r$ yields $\dot r=0$,
which is necessary for a circular orbit. From the minimum one finds
the energy of the test-particle as a function of angular momentum.
Obviously, one needs to keep $M$ and $M'$ constant during the
minimization, so the prescription to compute circular orbits becomes:
Minimize $\Eadm$ while keeping the angular momentum and the rest
masses constant.

These ideas have been formalized as variational principles for finding
equilibria for rotating and binary stars in Newtonian gravity. There
is also a similar variational principle for rotating stars in general
relativity \cite{Hartle:1967:VPF}. Binary systems in general
relativity are not strictly in equilibrium because they emit
gravitational waves. However, for orbits outside the innermost stable
circular orbit, the gravitational radiation reaction time scale is
much longer than the orbital period. It is therefore a good
approximation to treat the binary as an equilibrium system.

In this paper we apply this minimization principle to rotating binary
black hole systems.  Let the masses of the holes be $M_1$ and $M_2$,
the spins be $\ve S_1$ and $\ve S_2$, and the total angular momentum
of the system be $\ve J$.  We exploit the invariance under rescaling
of the mass by using dimensionless quantities $M_1/M_2$, $\ve
S_1/M_1^2$, $\ve S_2/M_2^2$, and $\ve J/\mu m$, where $m=M_1+M_2$ denotes
the total mass and $\mu=M_1M_2/m$ the reduced mass.  Then we adopt the
following straightforward prescription to locate quasi-circular
orbits: Minimize the scaled ADM energy $\Eadm/m$ with respect to the
separation of the holes, while keeping $M_1/M_2$, $\ve S_1/M_1^2$,
$\ve S_2/M_2^2$, and $\ve J/\mu m$ constant.

It is somewhat involved to carry out this simple prescription. The
computation of the ADM energy becomes more difficult than for the
Schwarzschild example above. More importantly, however, no rigorous
definitions exist for the mass or spin of an individual black hole in
a spacetime containing two black holes. We will address these issues
in Sec.~\ref{sec:Implementation}. Ultimately, we must use numerical
methods to generate and search among the solutions.  Our numerical
approach involves rootfinding, which is also described in
Sec.~\ref{sec:Implementation}.

In Sec.~\ref{sec:Results} we present the results of the effective
potential method. For the interpretation of these results, we need to
search for common apparent horizons in our binary black hole
data sets. These results are included in Sec.~\ref{sec:Results}, too.
We discuss our results and conclusions in Secs.~\ref{sec:Discussion}
and \ref{sec:conclusion}. The appendix contains details of the
apparent horizon searches.

\section{Implementation}\label{sec:Implementation}

In order to minimize the ADM energy while keeping $M_1/M_2$, $\ve
J/\mu m$, $\ve S_1/M_1^2$ and $\ve S_2/M_2^2$ constant, we need a
method to compute the ADM energy as a function of angular momentum,
masses and spins of the holes and separation. As a first step we
construct initial data $(\gamma_{ij},K_{ij})$ on a hypersurface as
described in \cite{Cook:1991:IDA,Cook:1993:TID,Cook:1994:TID}. Our
particular approach assumes conformal flatness of the 3-metric
$\gamma_{ij}$, maximal embedding of the hypersurface, as well as
inversion symmetry conditions on the 3-metric $\gamma_{ij}$ and on the
extrinsic curvature $K_{ij}$.  The effective potential method is
independent of these assumptions and works with all methods that compute
initial data. For example, in \cite{Baumgarte:2000:ISC}, the effective
potential method was used without assuming inversion symmetry.  In
particular, the assumptions of maximal embedding and conformal
flatness are not essential but merely convenient---maximal embedding
decouples the Hamiltonian and momentum constraints within the
initial-data formalism we use, and conformal flatness allows for an
analytic solution of the momentum constraints. One
disadvantage of conformal flatness is that Kerr black holes do not
admit conformally flat 3-metrics, at least for the simple time
slicings we are aware of.  In \cite{Garat:2000:NCF} it was shown that
the Kerr metric is not conformally flat at second order in the spin
parameter $S/M^2$.  Indeed, in Sec.~\ref{sec:PN} we identify this
deviation in our results.

Because we assume that the initial hypersurface is maximal, the
momentum and Hamiltonian constraints decouple. We follow the Bowen and
York \cite{Bowen:1980:TID} prescription to solve the momentum
constraint analytically.  Then we need only solve one
three-dimensional quasi-linear elliptic differential equation, the
Hamiltonian constraint.  It is solved on a so-called
{\v C}ade{\v z} grid using a multigrid
algorithm\cite{Cook:1993:TID}. The constructed data sets depend on
several input parameters, namely the radii and the positions of the
throats of the holes in the flat background space, $a_i$ and $\ve
C_i$, $i=1,2$, respectively, and their linear momenta and spins, $\ve
P_i$ and $\ve S_i$, $i=1,2$, respectively. We note that in this initial-data
prescription, $P_i$ and $S_i$ represent the \emph{physical} linear and
angular momentum of the black hole if it is isolated. We work in the zero
momentum frame, where $\ve P_2=-\ve P_1$, and choose $\ve P_i$
perpendicular to $\ve C_2-\ve C_1$ in order to realize a circular
orbit. Then the magnitude $P\equiv P_1=P_2$ is sufficient to describe
the linear momenta. Choosing $a_1$ as the fundamental length scale, we
are left with the following dimensionless input parameters: the ratio
of the throat radii $\alpha=a_1/a_2$, the dimensionless background
separation $\beta=|\ve C_1-\ve C_2|/a_1$, and the dimensionless linear
momentum and spins, $P/a_1$ and $\ve S_i/a_1^2$, $i=1,2$,
respectively.

From the initial data we can rigorously compute the ADM energy
$\Eadm$, the total angular momentum $\ve J$ and the proper
separation between the apparent horizons of each hole, $\ell$.  The
total angular momentum is evaluated as in Ref.~\cite{Cook:1994:TID}:
\begin{equation}\label{eqn:L_cook}
{\ve J} \equiv \left({\ve C}_1 - \ve O\right)\times{\bf P}_1
		+ \left({\ve C}_2 - \ve O\right)\times{\ve P}_2
		+ \ve S_1 + \ve S_2.
\end{equation}
Here $\ve O$ represents the point about which the angular momentum is
defined; it drops out immediately because $\ve P_1=-\ve P_2$.  When
orbiting black holes have spin, neither the individual spins of the
holes nor their orbital angular momentum $\ve L$ are rigorously
defined. We simply take $\ve L$ to be defined by
\begin{equation}
\label{eqn:L_def}
	\ve L \equiv \ve J - \ve S_1 - \ve S_2,
\end{equation}
with $\ve S_1$ and $\ve S_2$ defining the individual spins.

Finally, we need to define the masses of the individual holes. 
As in Ref.~\cite{Cook:1994:TID}, we define the mass of each hole
via the Christoudoulou formula:
\begin{equation}\label{eqn:M}
M_i^2=M_{ir,i}^2+\frac{S_i^2}{4M_{ir,i}^2},
\end{equation}
\begin{equation}\label{eqn:M_ir}
M_{ir,i}^2=\frac{A_i}{16\pi},
\end{equation}
where $A_i$ is the area of the event horizon of the $i^{\mbox{th}}$
hole.  Clearly this definition is only rigorous for a stationary
spacetime.  Moreover, we cannot locate the event horizon from the
initial data slice alone. Therefore we must resort to using the
apparent horizons areas in equations (\ref{eqn:M}) and
(\ref{eqn:M_ir}) instead. Apparent horizons can be determined from
initial data and in the present case their positions are known to
coincide with the throats of the holes \cite{Cook:1991:IDA}.
For a stationary spacetime, apparent horizons and event horizons
coincide, and in a general, well-behaved spacetime, the event horizon
must coincide with or lie outside of the apparent horizon.  In the
latter case we will underestimate the mass of the black hole by using
the apparent horizon area. Some of the results of this work indicate
that this happens for very small separations of the holes.

With the individual masses we can finally define the {\em effective
potential} as the non-dimensional binding energy of the system:
\begin{equation}\label{eqn:Eb}
\frac{E_b}\mu \equiv (\Eadm-M_1-M_2)/\mu.
\end{equation}
Since the mass-ratio $M_1/M_2$ is kept constant during the minimization,
minimizing $E_b/\mu$ is equivalent to minimizing $\Eadm/m$.

We construct initial data sets starting from the {\em input} parameters
$\alpha$, $\beta$, $P/a_1$ and $\ve S_i/a_1^2$, and compute the
{\em physical} parameters $E_b/\mu$, $M_1/M_2$, $J/\mu m$ and $\ve
S_i/M_i^2$. In order to construct an initial data set with certain
physical parameters we have to choose the input parameters
appropriately. This requires nonlinear rootfinding.

Within our effective potential approach, we will search for minima in
the binding energy as a function of the separation of the black holes.
Fortunately, it is not necessary to solve for a specific proper
separation $\ell/m$. It is sufficient to keep $\beta$ constant during
rootfinding and thus find a binary black hole configuration with some
separation $\ell/m$.  Our goal is to solve the following set of
equations [cf. Eqns.~(10a-d) of Ref.~\cite{Cook:1994:TID}]:
\begin{mathletters}
\label{eqn:roots}
\begin{eqnarray}
\label{eqn:X_root}
	\frac{M_1}{M_2}&=&\left[\frac{M_1}{M_2}\right]\\
\label{eqn:S1_root}
	\frac{S_1}{M_1^2}&=&\left[\frac{S_1}{M_1^2}\right]\\
\label{eqn:S2_root}
	\frac{S_2}{M_2^2}&=&\left[\frac{S_2}{M_2^2}\right]\\
\label{eqn:J_root}
	\frac{J}{\mu m}&=&\left[\frac{J}{\mu m}\right].
\end{eqnarray}
\end{mathletters}
The bracketed quantities on the right hand sides of
(\ref{eqn:X_root}-\ref{eqn:S2_root}) denote the physical values to be
reached, and the expressions on the left-hand side represent functions
of the background parameters $\alpha$, $P/a_1$, $S_1/a_1^2$ and
$S_2/a_1^2$ as well as the fixed $\beta$.

For non-rotating holes, equations (\ref{eqn:S1_root}) and
(\ref{eqn:S2_root}) are trivially satisfied by $S_1=S_2=0$. For
spinning holes this is no longer the case.  Hence, it seems one has to
solve the complete set of Eqns.~(\ref{eqn:X_root}--\ref{eqn:J_root}).
However, in any initial data scheme where the physical spins of the
black holes are directly parameterized, Eqns.~(\ref{eqn:S1_root}) and
(\ref{eqn:S2_root}) can be eliminated. First, we note again that if
the physical spins are directly parameterized, from
Eqn.~(\ref{eqn:L_def}) we find that we can replace rootfinding in
$J/\mu m$ by rootfinding in $L/\mu m$. Thus Eqn.~(\ref{eqn:J_root}) is
replaced by
\begin{equation}\label{eqn:L_root}
\frac{L}{\mu m}=\left[\frac{L}{\mu m}\right].
\end{equation}
In the zero momentum frame, Eqns.~(\ref{eqn:L_cook}) and
(\ref{eqn:L_def}) simplify to
\begin{equation}\label{eqn:L}
\frac{L}{a_1^2} = \beta \frac{P}{a_1}.
\end{equation}
Thus we can rewrite $S_1$ as
\begin{equation}\label{eqn:S1}
\frac{S_1}{a_1^2} =\frac{S_1}{M_1^2}\cdot\frac{M_1}{M_2}
	\cdot\frac{M_1M_2}{L}
	\cdot \beta \frac{P}{a_1}.
\end{equation}
For a solution of
Eqns.~(\ref{eqn:X_root}--\ref{eqn:S2_root},\ref{eqn:L_root}), the
first three terms on the right hand side of (\ref{eqn:S1}) take the
values of the desired physical parameters, so we can replace them by
these parameters throughout the rootfinding.  A similar result holds
for $S_2$.  We perform only two-dimensional rootfinding, in $\alpha$
and $P/a_1$, and set in each iteration
\begin{mathletters}
\begin{eqnarray}
\label{eqn:S1_soln}
	\frac{S_1}{a_1^2} &=& \left[\frac{S_1}{M_1^2}\right]
		\left[\frac{M_1}{M_2}\right]
		\left[\frac{L}{\mu m}\right]^{-1} \beta \frac{P}{a_1},\\
\label{eqn:S2_soln}
	\frac{S_2}{a_1^2} &=& \left[\frac{S_2}{M_2^2}\right]
		\left[\frac{M_1}{M_2}\right]^{-1}
		\left[\frac{L}{\mu m}\right]^{-1} \beta \frac{P}{a_1}.
\end{eqnarray}
\end{mathletters}

For an important subset of spin configurations, even one-dimensional
rootfinding is sufficient as can be seen as follows: Consider
equal-sized holes with equal spin magnitudes on both holes.  If both
spins are parallel to the orbital angular momentum, or both spins are
antiparallel, there exists a symmetry under exchange of the two
holes. Therefore $\alpha$ must be equal to $1$ and we are left with
one free parameter, $P/a_1$. If one spin is parallel to the orbital
angular momentum and the other spin is antiparallel, however, this
property is lost.  One hole is co-rotating with the orbital motion and
the other hole is counter-rotating. The choice $\alpha=1$ would result
in holes with slightly different masses. We thus need two-dimensional
rootfinding in $\alpha$ and $P/a_1$ for this case.

Each ``function evaluation'' for the rootfinding involves the
computation of an initial data set $(\gamma_{ij}, K_{ij})$. High
resolution solutions take between 30 minutes and several hours of CPU
time on one RS6000 processor.  For maximum efficiency, we first
perform rootfinding with a Newton-Raphson method\cite{Recipes} on low
resolution data sets.  The numerical values for $M_1/M_2$ and $J/\mu
m$ differ slightly between low resolution and high resolution
solutions, therefore we solve on low resolution for adjusted values of
$[M_1/M_2]$ and $[J/\mu m]$.  With the input parameters found in the
low resolution rootfinding, a high resolution computation is performed
to verify that equations (\ref{eqn:X_root}) and (\ref{eqn:J_root}) are
indeed satisfied at high resolution, and to adjust the offset used in
the next low resolution rootfinding. If necessary, this procedure is
repeated. On average each complete rootfinding takes fewer than two
high resolution computations.

Following our prescription, we now minimize the binding energy with
respect to separation while keeping $M_1/M_2$, $L/\mu m$ and $\ve
S_i/M_i^2$ constant.  The binding energy of a sequence of solutions with
these quantities held constant represents a contour of the effective
potential.  Our code starts at large separation $\beta$ and reduces
$\beta$ until a minimum in $E_b/\mu$ is bracketed.  Then the minimum
is located with Brent's method \cite{Recipes}, yielding a
quasi-circular orbit for the prescribed values of $J/\mu m$,
$M_1/M_2$, and $\ve S_i/M_i^2$.  Note that each computation of
$E_b/\mu$ during the minimization along an effective potential contour
requires rootfinding.

By computing quasi-circular orbits for different $J/\mu m$, but fixed
$M_1/M_2$ and $\ve S_i/M_i^2$, a {\em sequence} of quasi-circular
orbits is obtained. A binary black hole that radiates away energy and
angular momentum will follow such a sequence approximately, assuming
that the spin on each hole remains constant.  We step towards smaller
$J/\mu m$, and compute only as many points along each effective
potential contour as are required for the minimization.  As soon as we
do not find a minimum in the effective potential contours anymore we
expect to be beyond the innermost stable circular orbit.  We trace out
some complete effective potential contours around the last value of
$J/\mu m$ to check the behavior of these curves.

Finally, from the binding energy $E_b/\mu$ and the angular
momentum $J/\mu m$ along the sequence, we compute the orbital angular
frequency as
\begin{equation}\label{eqn:Omega}
\Omega=\left.\frac{\partial E_b}{\partial J}
\right|_{\mbox{\footnotesize{sequence}}}
\end{equation}

\section{Results}
\label{sec:Results}
The parameter space of spinning binary black holes is large -- one
can vary the mass ratio of the holes as well as spin directions and
magnitudes.  Astrophysically most interesting are holes that co-rotate
with the orbital motion, i.e.\ with both spins $\ve S_i$ parallel to
the orbital angular momentum $\ve L$. In addition to these co-rotating
configurations, we examine configurations with one co-rotating hole
and one counter-rotating hole, and configurations with two
counter-rotating holes. We have the following three families of
sequences:
\begin{itemize}
\item The ``\pps\ sequences'' with two co-rotating holes.
	\\ \mbox{}\hspace{-0.34in}\hrulefill\rule{0.4pt}{5pt}
\item The ``\pms\ sequences'' with one co-rotating and one counter-rotating
	 hole.
\item The ``\mms\ sequences'' with two counter-rotating holes.
\end{itemize}

We restrict ourselves to equal mass holes, $M_1=M_2\equiv M$ with
equal spin magnitudes $S_1=S_2\equiv S$.  As we will see, the
assumption of conformal flatness becomes questionable at high spins,
so we consider only spin magnitudes $S/M^2\le 0.50$.  We denote a spin
configuration by two plus or minus signs together with a number
specifying the spin magnitude on the holes. Thus ``$\pps0.25$''
denotes a configuration with two co-rotating holes and spin magnitudes
$S_1/M^2=S_2/M^2=0.25$.

\begin{figure}
\begin{picture}(240,240)
\put(0,0){\epsfxsize=3.25in\epsffile{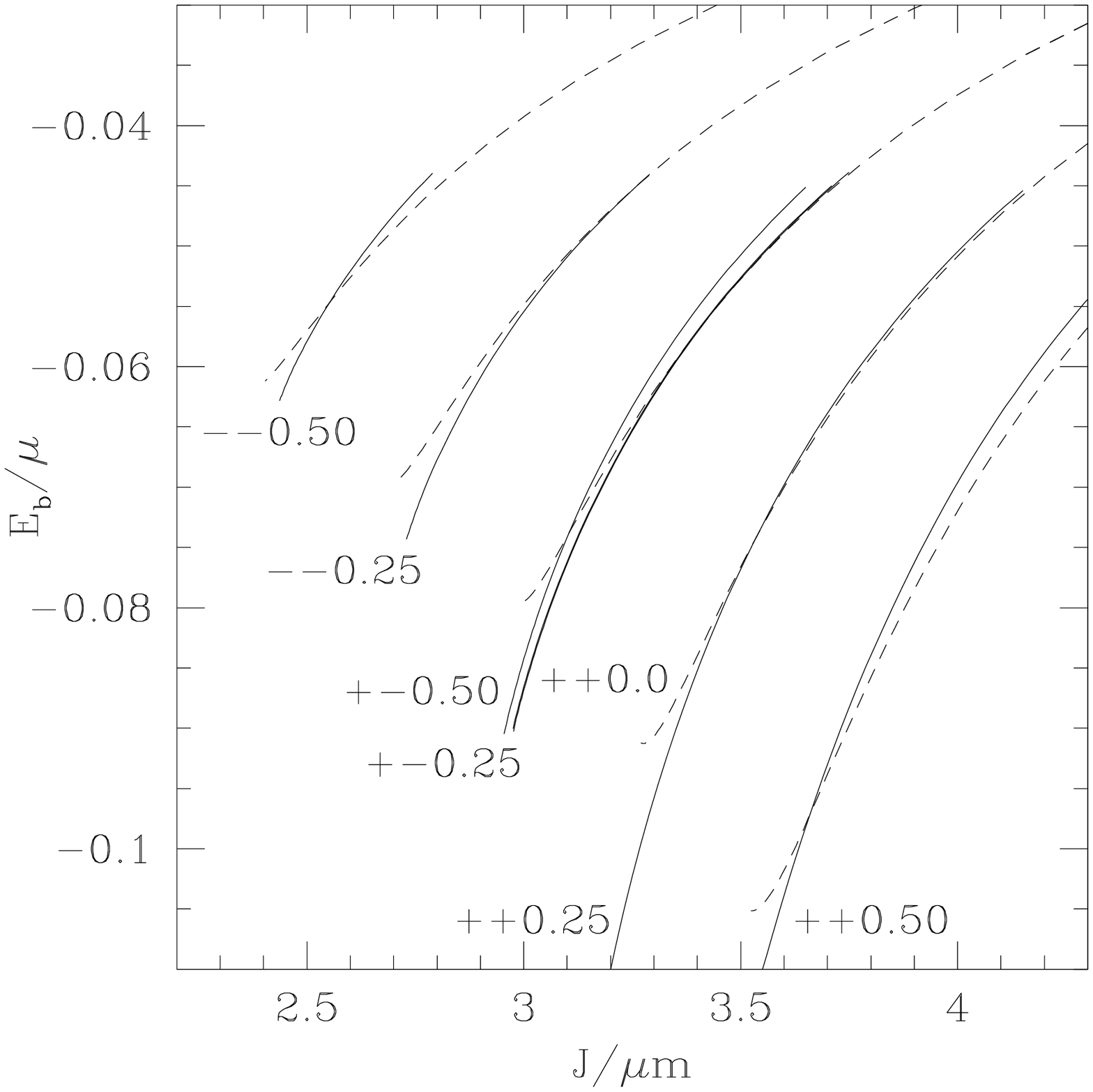}}
\end{picture}
\caption{\label{fig:seq_EbJ} Sequences of quasi-circular orbits for
different spin configurations.  Plotted is the binding energy
$E_b/\mu$ vs. the angular momentum $J/\mu m$ along the sequences.  The
solid lines represent the data, the dashed lines are the results based
on $\mbox{(post)}^2$-Newtonian theory.  As discussed later in this
paper, the effective potential method could not locate an ISCO for
the $\pps0.25$ and $\pps0.50$ sequences, although we believe each
sequence should terminate in one. }
\end{figure}

Quasi-circular orbits were computed for various values of $J/\mu m$
along each sequence. In Fig.~\ref{fig:seq_EbJ} the binding energy
$E_b/\mu$ along each sequence is plotted as a function of the angular
momentum $J/\mu m$. A binary black hole that loses energy and angular
momentum through gravitational radiation moves along such a sequence
if the spins of the individual holes remain constant.  The dashed
lines in Fig.~\ref{fig:seq_EbJ} represent the results of
$(\mbox{post})^2$-Newtonian theory which we describe in Sec.~\ref{sec:PN}.

Using equation (\ref{eqn:Omega}) we compute the orbital angular
frequency.  In Figs.~\ref{fig:seq_EbOmg} and \ref{fig:seq_JOmg}, the
binding energy and the angular momentum along the sequences are plotted
as a function of orbital frequency.

\subsection{Behavior at large separations}
\label{sec:PN}

\begin{figure}
\begin{picture}(240,240)
\put(0,0){\epsfxsize=3.25in\epsffile{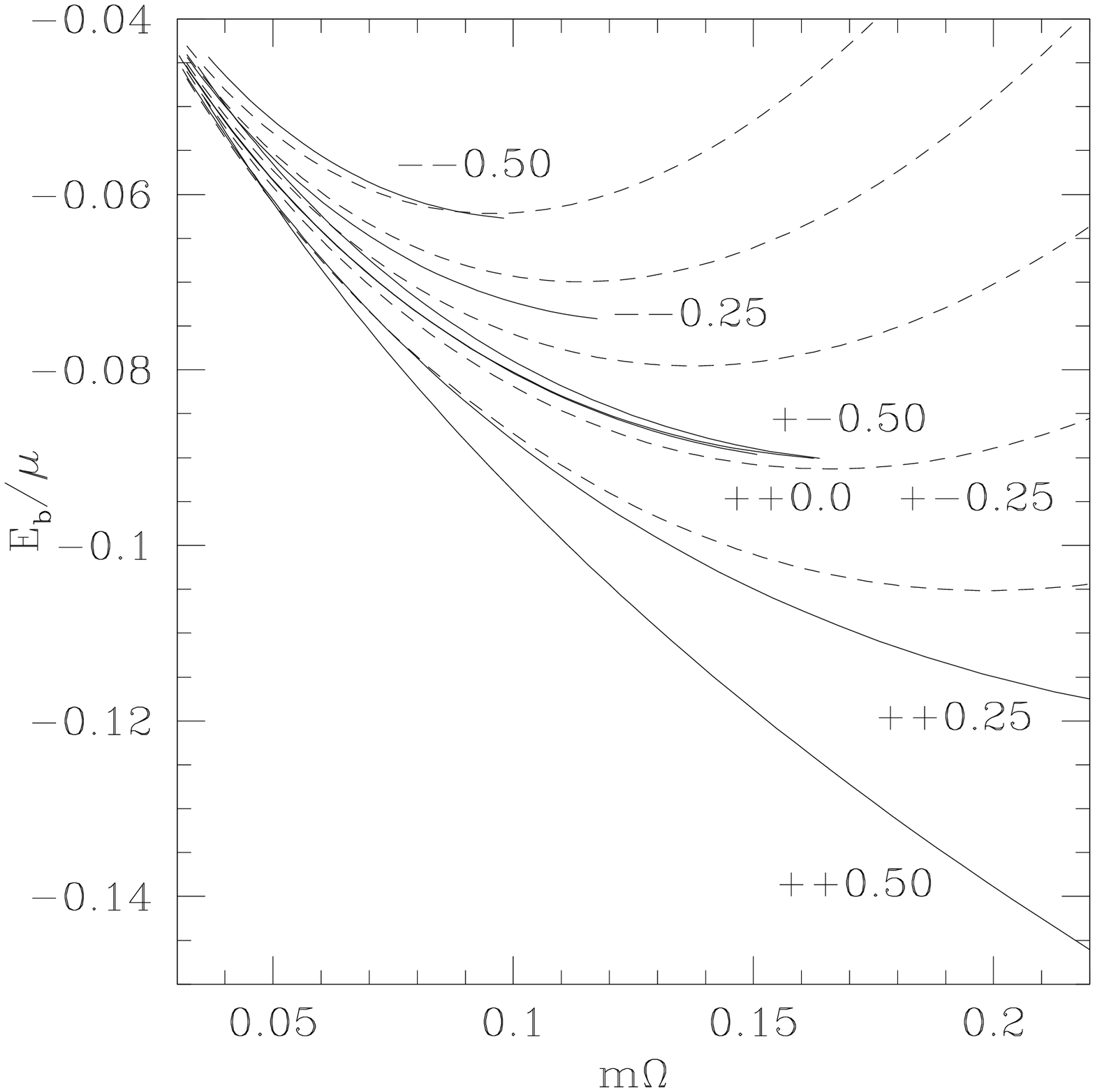}}
\end{picture}
\caption{\label{fig:seq_EbOmg} Sequences of quasi-circular orbits for
different spin configurations.  Plotted is the binding energy
$E_b/\mu$ vs. the orbital angular frequency $m\Omega$ along the sequences.
The solid lines represent the data, the dashed lines are the results
based on $\mbox{(post)}^2$-Newtonian theory.  As discussed later in this
paper, the effective potential method could not locate an ISCO for
the $\pps0.25$ and $\pps0.50$ sequences, although we believe each
sequence should terminate in one. }
\end{figure}

\begin{figure}
\begin{picture}(240,240)
\put(0,0){\epsfxsize=3.25in\epsffile{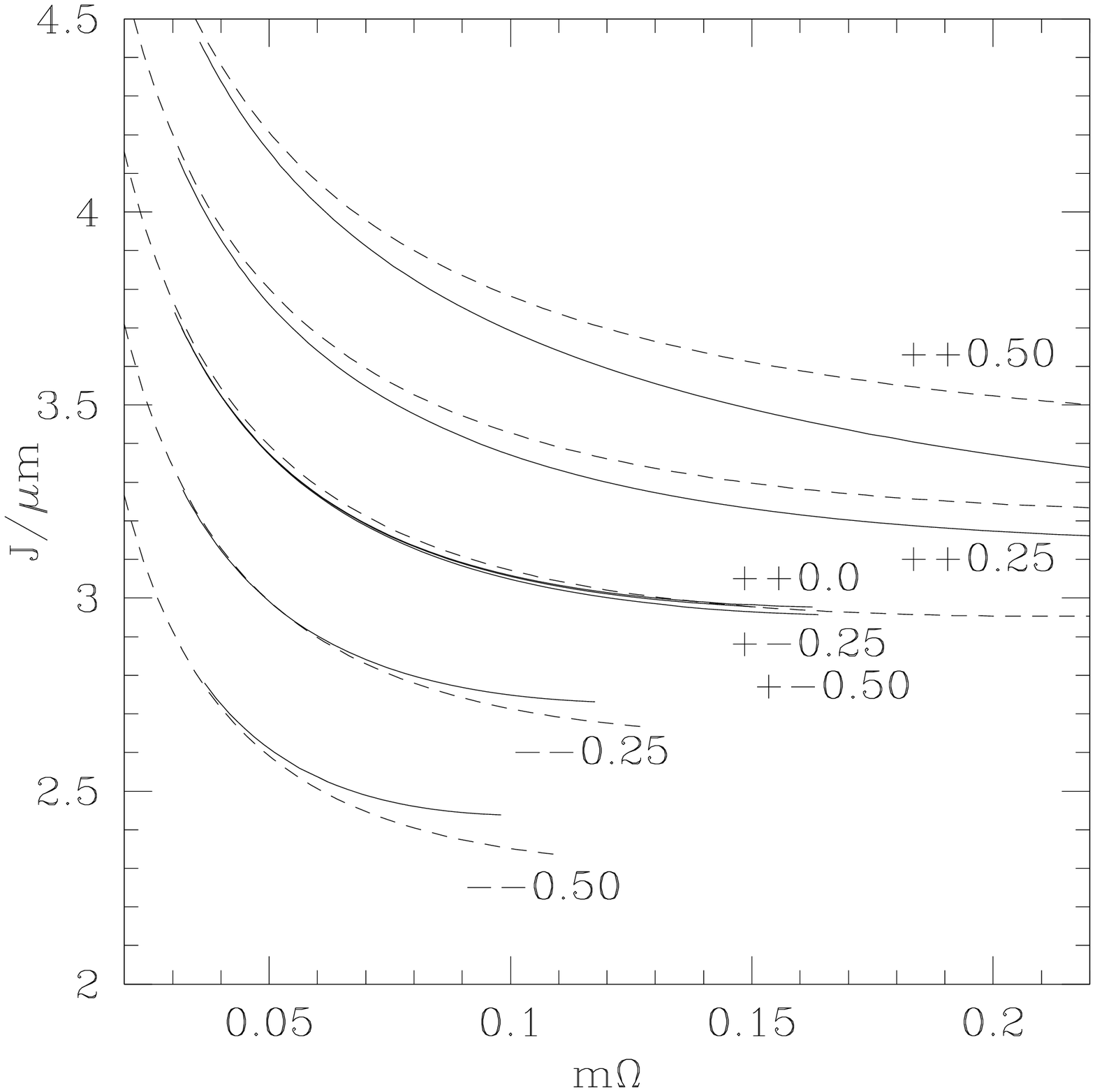}}
\end{picture}
\caption{\label{fig:seq_JOmg}
Sequences of quasi-circular orbits for
different spin configurations.  Plotted is the angular momentum $J/\mu m$
vs. the orbital angular frequency $m\Omega$ along the sequences.  The solid
lines represent the data, the dashed lines are the results based on
$\mbox{(post)}^2$-Newtonian theory.  As discussed later in this
paper, the effective potential method could not locate an ISCO for
the $\pps0.25$ and $\pps0.50$ sequences, although we believe each
sequence should terminate in one. }
\end{figure}

We compare our results to the $\mbox{(post)}^2$-Newtonian expansions
for spinning holes in quasi-circular orbit that were kindly provided
by L.\ Kidder.  The expressions for arbitrary spins and masses are
lengthy.  If one restricts attention to equal-mass holes, $M_1=M_2=M$,
$m=2M$, $\mu=M/2$, it turns out that only the {\em sum} of the spins
enters the $\mbox{(post)}^2$-Newtonian expansions.  In terms of
\begin{equation}\label{eqn:s_defn}
\ve s\equiv\frac{\ve S_1+\ve S_2}{M^2},
\end{equation}
and with $\hat\ve L$ being the unit-vector parallel to $\ve L$,
the $\mbox{(post)}^2$-Newtonian expansions become
\end{multicols}
\begin{mathletters}
\begin{eqnarray}
\frac{E_b}{\mu}\label{eqn:PN_Eb}
&=&-\frac{1}{2}\omg^{2/3}
\Bigg\{1-\frac{37}{48}\omg^{2/3}+\frac{7}{6}(\hat\ve L\cdot\ve s)\omg
-\left(\frac{1069}{384}+\frac{1}{8}
		\left[3(\hat\ve L\cdot\ve s)^2-\ve s^2\right]
	\right)\omg^{4/3}
\Bigg\}, \\
\left(\frac{J}{\mu m}\right)^2
\label{eqn:PN_J}
&=&\omg^{-2/3}
\Bigg\{1+2(\hat\ve L\cdot\ve s)\omg^{1/3}
+\left(\frac{37}{12}+\ve s^2\right)\omg^{2/3} \\
&&\hspace{1.87in}\mbox{}
+\frac{1}{6}(\hat\ve L\cdot\ve s)\omg
	+\left(\frac{143}{18}-\frac{37}{24}(\hat\ve L\cdot\ve s)^2
		-\frac{7}{8}\ve s^2\right)
\omg^{4/3}\Bigg\}.\nonumber
\end{eqnarray}
\end{mathletters}
\begin{multicols}{2}
\noindent
These expressions are plotted in
Figs.~\ref{fig:seq_EbJ}--\ref{fig:seq_JOmg} together with our results
from the effective potential method.  There is remarkable agreement.

The sum $\ve S_1+\ve S_2$ is zero for all \pms\ sequences with equal
spin magnitudes, so $\mbox{(post)}^2$-Newtonian theory predicts that
the \pms\ sequences are identical to the non-rotating sequence.  This
is remarkable, and indeed, in
Figs.~\ref{fig:seq_EbJ}--\ref{fig:seq_JOmg} the \pms\ sequences are
close to the $\pps0.0$ sequence.  However, a closer look reveals a
systematic behavior from which we can gain some insight into our
assumptions. For fixed angular momentum $J/\mu m$, consider the
difference in binding energy between a point on a \pms\ sequence
and a point on the non-rotating $0.0$ sequence,
\begin{equation}\label{eqn:DeltaE}
\Delta E_b/\mu(S)=\frac{E_b}{\mu}(\pms{S})-\frac{E_b}{\mu}(0).
\end{equation}

\begin{figure}
\begin{picture}(240,240)
\put(0,0){\epsfxsize=3.25in\epsffile{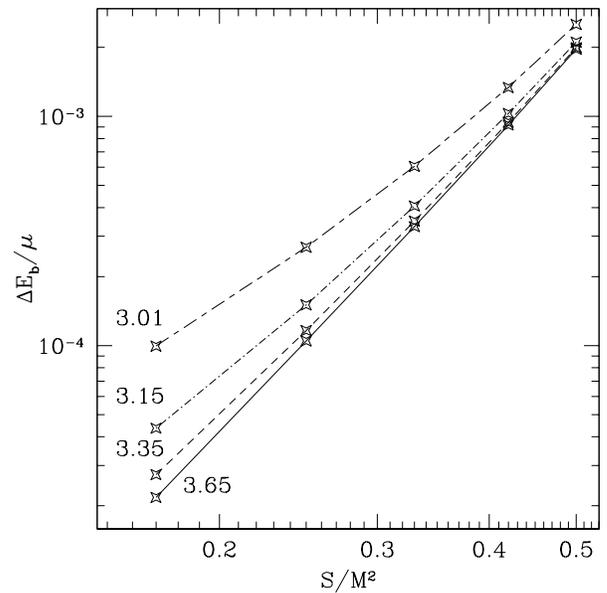}}
\end{picture}
\caption{\label{fig:spin4} Difference in binding energy $\Delta
E_b/\mu$ between \pms\ sequences and non-rotating sequence as a
function of spin of the \pms\ sequence for fixed angular momentum
$J/\mu m$.  Each curve is labeled by its value of $J/\mu m$.
$J/\mu m=3.01$ is very close to the ISCOs that have $J/\mu
m\approx 2.98$.  $J/\mu m=3.65$, $3.35$, $3.15$ and $3.01$ correspond to a
separation of $\ell/m \approx 12.3$, $9.6$, $7.7$ and $6.1$, respectively. }
\end{figure}

In Fig.~\ref{fig:spin4}, $\Delta E_b/\mu(S)$ is plotted as a function
of spin for several values of angular momentum $J/\mu m$.  $\Delta
E_b/\mu$ varies as the fourth power of spin.  This might be a physical
effect beyond $\mbox{(post)}^2$-Newtonian expansions, but for the
following reason it seems likely that one of our assumptions
introduces a non-physical contribution to $\Delta E_b/\mu$, too.
Figure~\ref{fig:spin4} strongly suggests that $\Delta E_b/\mu$ is
converging to a non-zero value as $J/\mu m$ (and thus separation)
increase, indicating
that there is a contribution to $\Delta E_b$ that is independent of
the separation of the holes.  For all spin configurations, $E_b$ must
approach zero in the limit of large separation, therefore any physical
contribution to $\Delta E_b$ should decrease with separation.
Moreover, a coupling between the holes, physical or unphysical, will
give rise to a separation-dependent contribution to $\Delta
E_b/\mu$. Therefore the separation-independent contribution must be a
non-physical effect due to properties of each {\em isolated} hole.  A
likely candidate is the underlying assumption of conformal flatness.
At large separations each hole should resemble a Kerr black hole,
which is {\em not} conformally flat.

Since the Kerr metric is the unique stationary state for a spinning
black hole, if the conformally flat initial data for a single hole
were evolved, the metric would relax to the Kerr metric and emit some
gravitational radiation.  Therefore the total energy contained in our
initial data slices is larger than in a more faithful conformally
non-flat data slice and $\Delta E_b/\mu$ should be positive, which it
indeed is.

We conclude that at large separations $\Delta E_b$ is contaminated by
an unphysical contribution because of the conformal flatness
assumption.  At small separation there might be additional physical
contributions beyond the $(\mbox{post})^2$-Newtonian order.

\subsection{Behavior at small separations -- ISCO}
\label{sec:results:ISCO}

In this section we report the key results of this work -- the
spin dependence of the innermost stable circular orbit. As we will
see, the interpretation of our data at small separations is somewhat
complicated. At large separations, the assumptions and approximations
we have used are reasonable, except for the assumption of conformal
flatness when the holes are spinning.  At small separations, the
interaction between the two black holes becomes relatively strong,
and our approximations begin to break down. Near the ISCO, we must
evaluate the quality of our assumptions to determine how reliable our
results are.

\begin{figure}
\begin{picture}(240,240)
\put(0,0){\epsfxsize=3.25in\epsffile{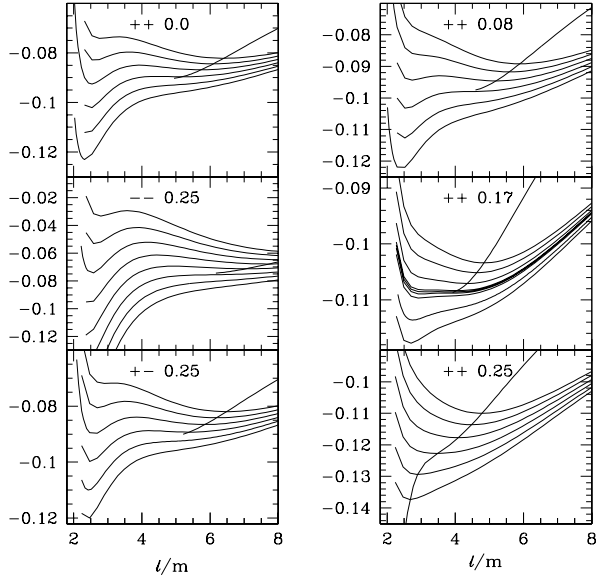}}
\end{picture}
\caption{\label{fig:Eb_seq}
Constant $J/\mu{m}$ contours of the effective potential $E_b/\mu$ as a
function of separation $\ell/m$ for various spin configurations. The
curves are spaced in steps of $\Delta{J}/\mu{m}=0.02$ except for the
$\mms0.25$ and the $\pps0.17$ configurations, which have steps of
$0.04$ and $0.01$, respectively. Also plotted is the sequence of
quasi-circular orbits connecting the minima of the effective potential. }
\end{figure}

In the neighborhood of each tentative ISCO, we compute a set of
complete effective potential contours. These are shown in
Fig.~\ref{fig:Eb_seq}. In each plot, the binding energy $E_b/\mu$
is shown as a function of separation $\ell/m$ for several different
values of angular momentum $J/\mu m$. Also plotted is the sequence of
quasi-circular orbits passing through the minima of the effective
potential.  Figure~\ref{fig:Eb_seq} shows the non-rotating sequence
$\pps0.0$, one example each of a \mms\ and a \pms\ sequence, and three
\pps\ sequences with different spin magnitudes.

Examining the constant $J$ contours of the effective potential for
{\em fixed} spin configurations, we find that they fall into three
regimes separated by critical values that we will label $J_A$ and
$J_B$. Contours with $J>J_A$ exhibit a single minimum positioned at
large separation $\ell/m$. This minimum moves inward as the angular
momentum decreases, i.e.\ the holes approach each other as angular
momentum and energy are radiated away.  We call this the ``outer''
minimum. As $J$ passes through the critical value $J_A$, a new
``inner'' minimum appears inside the outer minimum. In this region,
contours of the effective potential have two minima separated by a
local maximum. The maximum corresponds to the well known unstable
circular orbit of a Schwarzschild black hole. As $J$ decreases
further, $J_A>J>J_B$, the maximum moves outward whereas the outer
minimum continues to move inward -- the quasi-circular orbit associated
with the outer minimum continues to shrink. As $J$ passes through the
second critical value $J_B$ the outer minimum and the maximum meet in
an inflection point and disappear. The quasi-circular orbit associated
with the outer minimum disappears and this inflection point is
identified with the ISCO.  For $J<J_B$, only the inner minimum
remains.

This behavior for the non-rotating sequence was already found in
\cite{Cook:1994:TID}. There, the inner minimum was dismissed as
unphysical, since the underlying assumptions become weaker at small
separations of the holes, and since a common event horizon might form.
We will discuss this ``unphysical'' region and the possibility and
consequences of the formation of a common event horizon below.
But first we continue discussing the behavior of the effective
potential for different spin configurations. 

As we increase the spin magnitude for the \mms\ configurations, the two
critical angular momentum values $J_A$ and $J_B$ move away from each
other. We see a more pronounced local maximum and the $E_b$ curves
look similar to the effective potential of Schwarzschild for a larger
interval of angular momenta. The ISCO moves outward to larger
separations as spin increases.

\begin{figure}
\begin{picture}(240,240)
\put(0,0){\epsfxsize=3.25in\epsffile{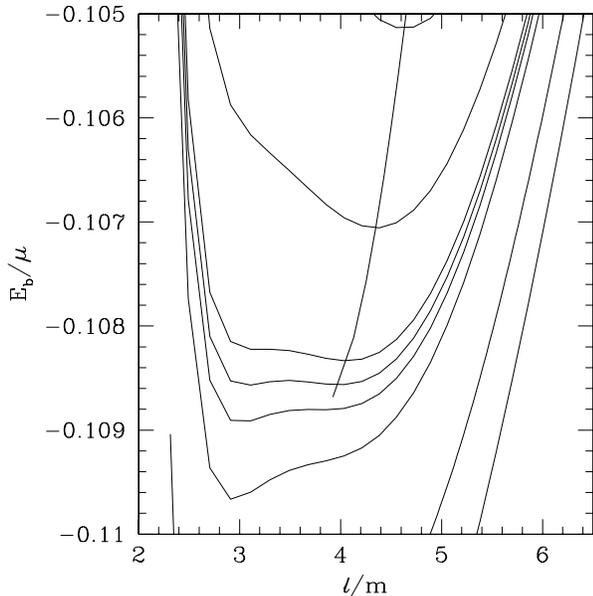}}
\end{picture}
\caption{\label{fig:PP017enlarge}
Enlargement of the $\pps0.17$ sequence of Fig.~\ref{fig:Eb_seq}.  The
displayed effective potential contours (top to bottom) correspond to
angular momenta $J/\mu m=3.12$, $3.11$, $3.104$, $3.103$, $3.102$,
$3.10$, $3.09$ and $3.08$. Also shown is the sequence of
quasi-circular orbits.}
\end{figure}

Conversely, as we increase the spin magnitude for the \pps\
configurations the interval $(J_B, J_A)$, where two minima and a local
maximum exist becomes smaller.  Slightly above $S/M^2=0.17$, $J_A$ and
$J_B$ merge and for $S/M^2\gtrsim0.17$, the regime with two minima and a
maximum is {\em not} present.  Figure~\ref{fig:PP017enlarge}
illustrates the small interval $(J_B, J_A)$ with an enlargement of the
$\pps0.17$ sequence.  As long as the regime with two minima and a
maximum is present, we can still define the ISCO by the inflection
point. It moves towards smaller separation of the holes as the spin is
increased.  However, since the inflection point ceases to exist at
some spin magnitude, we cannot define an ISCO for all $S/M^2$.
Therefore the \pps\ sequences displayed in
Figs.~\ref{fig:seq_EbJ}--\ref{fig:seq_JOmg} do not terminate.
Furthermore, we need a more careful analysis to determine whether the
ISCO properties for spin magnitudes close to the critical value
$S/M^2\approx 0.17$ are reliable.

The \pms\ configurations are very similar to the non-rotating
one. Given the weak dependence on spin within the \pms\ sequences, this
is not surprising.  We do not consider the \pms\ configurations
further.

\begin{figure}
\begin{picture}(240,240)
\put(0,0){\epsfxsize=3.25in\epsffile{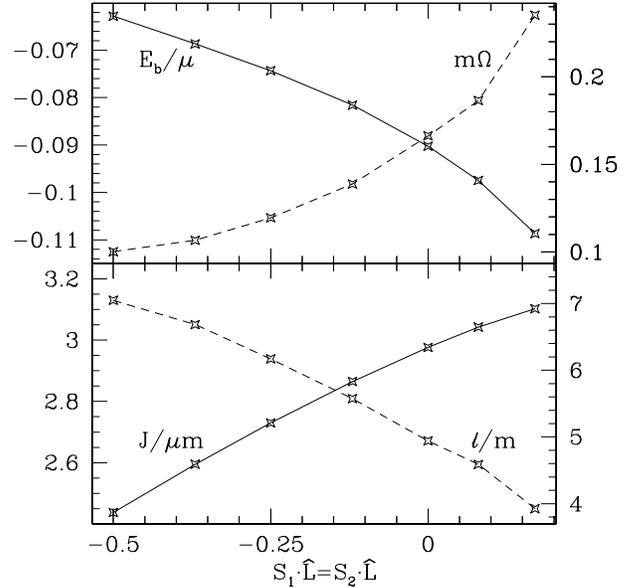}}
\end{picture}
\caption{\label{fig:ISCO}
Values of several physical parameters at the ISCO of the \pps\ and
\mms\ sequences.  Plotted are the binding energy $E_b/\mu$, the
orbital angular frequency $m\Omega$, the total angular momentum $J/\mu
m$ and the proper separation between the holes, $\ell/m$ as a function
of spin $S/M^2$ on the holes. The \pps\ sequences are plotted along
the positive part of the horizontal axis, the \mms\ sequences along
the negative part as $-S/M^2$. The vertical axes on the left side
belong to $E_b/\mu$ and $J/\mu m$. }
\end{figure}

Figure~\ref{fig:ISCO} and Table~\ref{tab:ISCO} summarize the orbital
parameters at the ISCO as a function of spin for the \mms\ sequences
and the \pps\ sequences.  The numerical errors in $E_b/\mu$, $L/\mu m$ and
$J/\mu m$ are less than 1 per cent, while $m\Omega$ and $\ell/m$ are
accurate to a few percent. However, for the \pps\ sequences the
systematic errors of our approach might be much larger.  The table
also includes ISCO parameters for a test mass orbiting a Kerr black
hole obtained from formulas in \cite{Bardeen:1972:RBH}.

\subsection{Common apparent horizons}

A common event horizon might be responsible for the strange
behavior of the effective potential at small separations, because once
a common event horizon forms, there are no longer two distinct
black holes.  It would be helpful to know the critical separation
where a common event horizon first forms.  However, in order to
locate the event horizon, knowledge of the complete spacetime is
needed. In the present case, only data on one time-slice is available,
and so we can only search for common apparent horizons. Since
the event horizon must lie outside the apparent horizon, the formation
of a common apparent horizon places a firm bound on the
formation of an event horizon.

Searches for a common apparent horizon were carried out for
several spin configurations.  Details of the apparent horizon finder
and the method used to discern the formation of a common
apparent horizon are given in the Appendix.  In Table~\ref{tab:AH},
the results of the apparent horizon searches are listed.

For fixed spin configurations the common apparent horizon forms at
larger separation for larger angular momentum.  This can be seen from
the $\mms0.25$ and $\pps0.0$ sequences.  For varying spins and angular
momentum close to the ISCO values, the proper separation between the
throats at the formation of the common apparent horizon depends weakly
on the spin.  It decreases from $\ell/m\approx 2.3$ for the $\mms0.37$
sequence down to $\ell/m\approx 2.0$ for the $\pps0.17$ sequence.

{\em Notice that the segment of parameter space where common apparent
horizons form does {\em not} include the sequence of quasi-circular
orbit configurations.}  Indeed, the common apparent horizons form at a
separation inside the inner minimum where the effective potential {\em
increases} with decreasing separation.

\begin{figure}
\begin{picture}(240,240)
\put(0,0){\epsfxsize=3.25in\epsffile{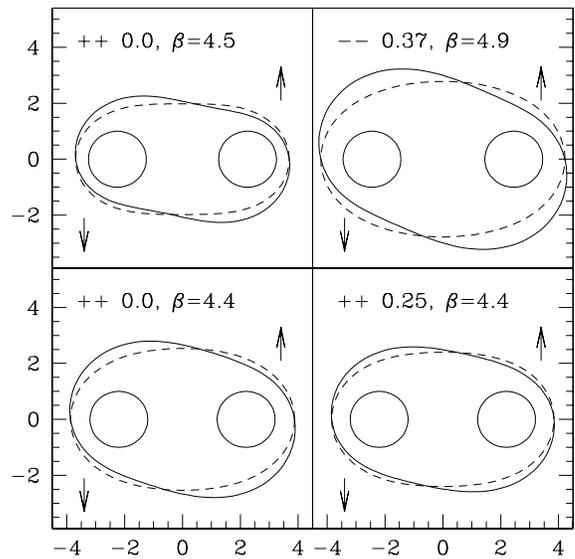}}
\end{picture}
\caption{\label{fig:AHshapes}
Shapes of the common apparent horizons for different spin
configurations. Circles denote the throats of the holes. The solid
lines are cuts in the plane of orbital motion (arrows indicating the
direction of motion), the dashed lines represent cuts normal to the
plane of motion.  }
\end{figure}

The search for the onset of common apparent horizons also provides
the actual surfaces. In Fig.~\ref{fig:AHshapes} some apparent
horizon surfaces just inside the formation of a common apparent
horizon are plotted.  The circles represent the throats of the
holes. The solid lines represent a cut through the plane of orbital
motion of the holes, arrows indicating the direction of linear
momentum of the holes.  The dashed lines are cuts through the plane
perpendicular to the plane of motion and parallel to the spins of the
holes.  We find that the apparent horizons lag behind the orbital
motion, with the amount of lag being larger for counter-rotating than
for co-rotating holes.

\section{Discussion}
\label{sec:Discussion}

We found that the effective potential contours at very small separation
{\em increase} with decreasing separation. This is in contrast to the
usual shape of the effective potential for a Schwarzschild or a Kerr
black hole, which tends to $-\infty$ at sufficiently small
separations.

This behavior can be interpreted in the light of the common
apparent horizon searches. The common apparent horizon that was
found to form at a small separation of the holes might influence the
observed effective potential as follows: The event horizon must
lie outside the apparent horizon.  Therefore a common event
horizon must form before a common apparent horizon forms.  To
accomplish this the event horizons around the individual holes must
grow towards this common event horizon.  Thus, even before
formation of a common event horizon, the individual event horizons
will no longer be close to the individual apparent horizons and the
areas of the event horizons of the individual holes must be larger
than the areas of their apparent horizons.  Therefore, equations
(\ref{eqn:M}) and (\ref{eqn:M_ir}) will {\em under}-estimate the mass
of the holes. We denote this underestimate by $\Delta M$.  Consider
the effect this underestimate of $M$ has on the binding energy.  
The numerator of (\ref{eqn:Eb}) will be {\em over}-estimated by a 
relative amount of
\begin{equation}\label{eqn:Eb2}
\frac{2\Delta M}{|\Eadm-2M|}
=\frac{4}{|E_b/\mu|}\frac{\Delta M}{M}
\gg \frac{\Delta M}{M}.
\end{equation}
At the same time, the denominator of (\ref{eqn:Eb}) and the
denominator of the scaled angular momentum (\ref{eqn:J_root}) change
too, leading to an underestimate of the binding energy $E_b/\mu$.
However, the relative changes of these denominators are only of the
order of $\Delta M/M$, so that the overestimate from
Eqn.~(\ref{eqn:Eb2}) dominates. It might
well be that this overestimate is so large that it counter-balances
the decreasing effective potential that one might expect in analogy to
Schwarzschild or Kerr black holes.

This idea leads to the following picture to explain the observed
effective potential curves: At large separation of the holes, the
masses of the holes and the effective potential are reliable and we
see an effective potential that looks similar to a Schwarzschild black
hole. Consider, for example, the $\pps0.0$ sequence: For $J$ slightly
above its ISCO value we see the (outer) minimum of the stable
quasi-circular orbit and a maximum corresponding to an unstable
circular orbit. As $J$ increases, the stable circular orbit moves
outwards and the unstable one moves inwards. Once the maximum
corresponding to the unstable orbit moves too far in, the $\Delta M/M$
contamination of the effective potential ``eats up'' the maximum and
it disappears.

Now we turn on spin.  We found that a common apparent horizon forms at
approximately the same proper separation, independent of the spin of
the holes.  It seems reasonable that the $\Delta M/M$ error is also
weakly dependent on the spin, and also the separation of the holes,
where $\Delta M/M$ becomes significant.  For the \mms\ sequences the
ISCO moves to larger separations.  Thus the maximum in the effective
potential (the unstable orbit) will survive for a larger range of
separations and angular momenta $J$.  Conversely, for the \pps\
sequences, the ISCO moves inwards, closer to the separation where
$\Delta M/M$ becomes significant. The maximum in $E_b/\mu$ is lost
almost immediately, and in the extreme limit of $S/M^2>0.17$, it does
not show up at all.

This scenario is sufficient to capture the complete behavior of the
effective potential as a function of $J$ and spin.  What does this
picture imply for the validity of our ISCO results from
Table~\ref{tab:ISCO}?  We expect that $\Delta M/M$ decays rapidly with
increasing separation, so the ISCO data for the non-rotating sequence
$\pps0.0$ as well as for the \mms\ sequences should be sound.
However, because $\Delta M$ changes the characteristic behavior for
the \pps\ configurations even for $S/M^2<0.17$, the \pps\ sequences will
be affected. Let us consider how these changes affect our estimates
of circular orbits.

\begin{figure}
\begin{picture}(240,240)
\put(0,0){\epsfxsize=3.25in\epsffile{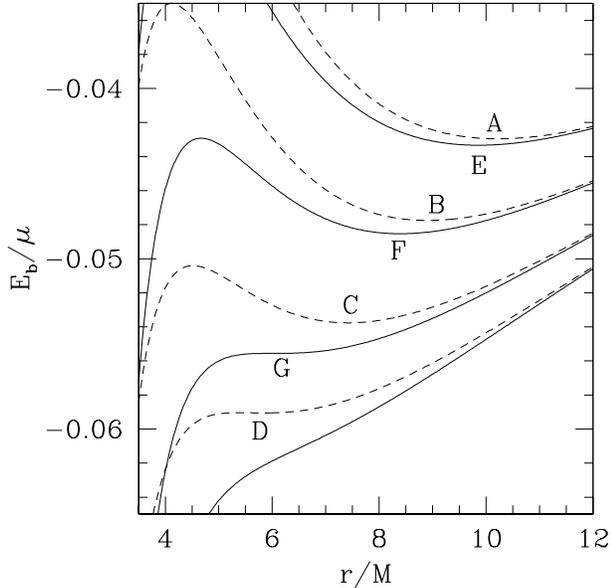}}
\end{picture}
\caption{\label{fig:toy}
Illustration of the effects of a systematic underestimation of
$E_b/\mu$. The dashed lines represent the observed effective potential
contours for some values of $J$. The points A, B, and C correspond to
circular orbits.  The ISCO is at D.  Assuming that the true binding
energy is smaller, with the deviation increasing as the separation
decreases, yields true effective potential contours similar to the
solid lines. The true circular orbits are at E and F and the true ISCO
is at G.  We find that the minima of the true contours will
lie at smaller separation (for the same $J$).  The angular frequency is
given by $\Omega=dE_b/dJ$. Using the points A and B, we see that the
observed $dE_b$ is smaller than the true one, so we under-estimate
$m\Omega$.  For fixed $J$, true circular orbits will occur at smaller
separation, but the true ISCO will appear at larger $J$ than we have
observed.  These effects counteract each other, making it impossible
to predict their effect on the true ISCO.  }
\end{figure}

Figure~\ref{fig:toy} illustrates the effect of the $\Delta M/M$
contamination on the effective potential contours. As we noted above,
the $\Delta M/M$ contaminations of the binding energy overestimates
the binding energy of an effective potential contour.  Since this
error increases as the separation decreases, our estimates for the
separation at a given value of angular momentum are also too high, and
our estimates of the orbital angular velocity $m\Omega$ are too
low. Unfortunately, we cannot determine whether our estimates for the
location of the ISCO are too high or too low.  While our estimates for
the separation of a given orbit are too high, we see that the true
ISCO will occur at a larger value of the total angular momentum than
we estimate.  These effects oppose each other.

The angular momentum at the ISCO, $J/\mu m$, increases with spin for the
\pps\ configurations. It is interesting to examine whether the final
black hole resulting from a merger of such a spinning binary black
hole can violate the Kerr limit on spin of a black hole. From
(\ref{eqn:M}) we find 
\begin{equation}
M_{ir}^2=\frac{M^2}{2}\left(1+\sqrt{1-\frac{S^2}{M^4}}\right).
\end{equation}

By the area theorem, the final irreducible mass must satisfy
$M_{ir,f}^2\ge2M_{ir}^2$, where equal mass holes were assumed. The
final angular momentum cannot exceed the angular momentum at the ISCO,
$J_f\le J$. With these two constraints and by virtue of the
Christoudoulou formula (\ref{eqn:M}), we find

\begin{equation}
\frac{M_f^2}{M_{ir,f}^2}\le 1+\frac{\left(J/\mu m\right)^2}{4\left(
1+\sqrt{1-\left(S/M^2\right)^2}\;\right)^2}
\end{equation}

A Kerr black hole has always $M^2/M_{ir}^2\le 2$ with equality in the
extreme Kerr limit. With data from Table~\ref{tab:ISCO} we find for
the $\mms0.50$ sequence $M_f^2/M_{ir,f}^2\le 1.43$ and for the $\pps0.17$
sequence $M_f^2/M_{ir,f}^2\le 1.61$. These values correspond to spin
parameters of $J/M_f^2\le 0.92$ and $J/M_f^2\le 0.97$, respectively.
Hence the merged black hole might be close to the Kerr limit, but will
not violate it.

\subsection{\pms\ Sequences and conformal flatness}

The $(\mbox{spin})^4$ effect illustrated in Fig.~\ref{fig:spin4}
suggests that the assumption of conformal flatness might lead to
inaccurate results.  This is particularly important for analysis of
gravitational waves.  As seen in Fig.~\ref{fig:spin4}, for spinning
holes with $S/M^2\sim 0.50$ the assumption of conformal flatness
results in an unphysical gravitational wave content of the order
of $\sim 2\cdot 10^{-3}\mu\sim 5\cdot 10^{-4}m$. This is less than
$0.1$ percent of the total mass and a few percent of the binding
energy $E_b$.  If the gravitational energy radiated away is less than
1\% of the total mass, then the gravitational wave content due to an
unsuitable initial data slice is a significant contamination.

\section{Conclusion}
\label{sec:conclusion}

In this work, we have constructed sequences of quasi-circular orbits
for equal-sized, spinning black holes.  At large separations, the
results we have obtained match well with $(\mbox{post})^2$-Newtonian
expansions, although there is a clear contamination of the data because of
the assumption of conformal flatness. The main results of this paper,
displayed in Table~\ref{tab:ISCO} and Fig.~\ref{fig:ISCO}, reveal the
behavior of the ISCO for the cases where the spins of the holes are
either both co-rotating (\pps) or counter-rotating (\mms) with respect
to the orbital motion. For co-rotation, the ISCO moves inwards with
increasing spin and the orbital angular frequency increases. For
counter-rotation the ISCO moves outward and the orbital angular
frequency decreases.  In fact, we find that the orbital angular
frequency changes by almost a factor of 2 between the $\mms0.50$
sequence and the $\pps0.08$ sequence. We have noted a systematic error
in our results that has its origins in an underestimation of the mass
of each black hole when they are close together.  For the ISCO, this
implies that our results are most accurate (ignoring the errors due to
conformal flatness) when the holes have large counter-rotating
spins, and the error increases as we move to configurations with large
co-rotating spins.  In fact, the error becomes so large in the \pps\
sequences that our method cannot locate the ISCO when
$S/M^2\gtrsim0.17$.

Our results clearly show the need to give up the simplifying
assumption of conformal flatness if we are to construct
astrophysically realistic black hole initial data.  This is certainly
not a new realization, but this is the first time that the effects of
the conformal flatness assumption have been seen so clearly in the
context of black hole binaries.  Work toward more astrophysically
realistic initial data has begun \cite{Matzner:1999:IDC}. This
improvement in the initial data is needed for all separations. It
remains to be seen what impact this improvement will have on the
process of locating quasi-circular orbits when the holes are close
together. It is likely that the systematic underestimate of the mass
will still be significant.  If so, an improved method for locating
quasi-circular orbits and the ISCO will be useful.

\begin{acknowledgements}

We thank Larry Kidder and Mark Scheel for helpful discussions.
This work was supported in part by NSF grants PHY-9800737 and
PHY-9900672 and NASA Grant NAG5-7264 to Cornell University, and NSF
grant PHY-9988581 to Wake Forest University. Computations were
performed on the IBM SP2 at Cornell Theory Center, and on the Wake
Forest University Department of Physics IBM SP2 with support from an
IBM SUR grant.

\end{acknowledgements}

\appendix
\section*{Common apparent horizons}
\label{sec:AHfinder}
Here we provide details of the apparent horizon (AH) finder.  We use
the AH finder described in \cite{Baumgarte:1996:IAF}.  The AH surface
is expanded in spherical harmonics up to some order $L$.  The
apparent horizon, as a marginally outer trapped surface, has
everywhere vanishing expansion and is located by minimizing the square
of the expansion over the surface.  We use convergence with
increasing expansion order $L$ to diagnose the formation of a
common AH. Therefore high expansion orders $L$ are needed as
well as reliable convergence of the minimization routine to the true
minimum of the square of the expansion.

The Powell minimization used in \cite{Baumgarte:1996:IAF} is too slow
for high-order expansions. We replaced it by a DFP method with finite
difference approximations of the Jacobian \cite{Recipes}. For the
modest expansion order $L=6$, DFP is already ten times faster than
Powell's method.

Furthermore, we take advantage of the symmetries of the AH
surface. The holes are located along the $\hat z$-axis at $z=\pm
\beta/2$. Their linear momenta point in the $\pm \hat x$-direction and
the spins are directed along the $\pm \hat y$-axis. It is
straightforward to show that these choices imply that the AH surface
is invariant under reflection at the $xz$-plane, $y\to -y$. This
symmetry constrains the coefficients $A_{lm}$ of the expansion in
spherical harmonics to be real. Moreover, for the \pps\ and \mms\ 
configurations with equal sized holes and equal spin magnitudes, the
configuration is symmetric under rotation by $180^\circ$ around the
$\hat y$-axis, this is $(x,y,z)\to (-x, y, -z)$. Both symmetries
together force $A_{lm}=0$ for odd $l$ and $A_{lm}$ to be real for even
$l$.  Hence the number of free parameters in the minimization routine
can be reduced by almost a factor of four.

To prevent convergence to spurious local minima, it is vital that the
function that is minimized be as smooth as possible.  Therefore we use
second order spline interpolation to provide the required data for the
AH finder.  Compared to bicubic interpolation, the
spline interpolation somewhat decreased the number of iterations
needed in the minimization routine, but more importantly it
significantly reduced the probability of getting stuck in a local
minimum.  In addition, many rays were used to reduce the anisotropies
introduced by the discrete position of the rays. Finally, we
distribute the rays {\em non}-uniformly in solid angle. The reason for
this is simple: The common AH surface will be very oblate along
the $\hat z$-axis, since it must encompass the two throats located
along the $\hat z$-axis. The polar regions of the AH surface are close
to the throats and the conformal factor changes rapidly. These regions
are particular important, but the standard distribution uniform in
$\cos\theta$ places relatively few rays in the polar regions.
Therefore we implemented a procedure that distributes the rays in
proportion to an arbitrary ray-density function $f(\theta)$. A
uniform distribution of rays is represented by
$f(\theta)=\mbox{const.}$, whereas we used $f(\theta)=1+\cos^2\theta$,
resulting in a doubled density of rays close to the poles.

With the improved AH finder, we performed extensive tests with various
numbers of rays. As a rule of thumb, about ten times more rays as free
minimization parameters are necessary to ensure reliable convergence
to the true minimum of the square-sum of the expansion.

We used expansions up to order $L=16$ and up to 64x48 rays (64 in
$\theta$ direction, 48 in $\phi$).  We perform a set of AH searches,
starting at $L=2$ and increasing $L$ by $2$ between searches. The
result of the previous search is used as the initial guess for the
next higher expansion order.  Such a set of expansions from $L=2$ to
$L=16$ takes typically about 2 hours CPU time on a RS6000 processor.

A disadvantage of an AH finder based on a minimization routine is
that the minimization routine will {\em always} find a minimum.  It
does not matter whether there actually is a ``true'' apparent horizon,
or whether there is only a surface with small but non-zero
expansion. And even for a true AH, the result of the minimization will
be non-zero because of the finite grid resolution in the underlying
elliptic solver and finite expansion order in spherical
harmonics. Therefore we need a method to discern a ``true'' AH from a
mere minimum in the square of the expansion.

For a true AH, the square of the expansion is exactly zero, therefore
we expect that the residual of the minimization tends to zero as the
resolution of the elliptic solver and the expansion order $L$ are
increased. With increasing $L$, the error in the approximation of the
surface by spherical harmonics should decrease {\em exponentially}. On
the other hand, for a mere minimum in the expansion, we expect that
the residual of the minimization tends towards a {\em non-zero} limit
as the resolution of the elliptic solver and the expansion order $L$
is increased. We use this signature to discern the formation of a
common apparent horizon.

\begin{figure}
\begin{picture}(240,240)
\put(0,0){\epsfxsize=3.25in\epsffile{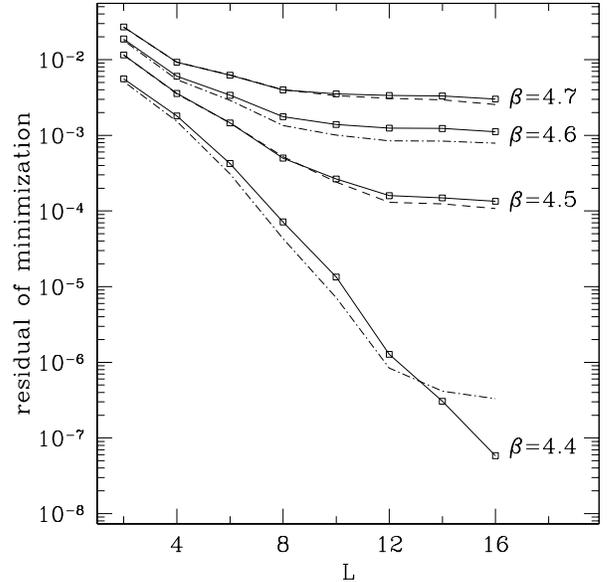}}
\end{picture}
\caption{\label{fig:AHconv}
Residual of the minimization in the AH finder as a function of
expansion order L. The number of rays used was $N_\theta=64$,
$N_\phi=48$.  The different solid lines represent different
separations of the holes along an effective potential contour with
$J/\mu m=3.29$ on the $\pps0.25$ sequence.  The dashed lines are the
results of minimizations with $N_\theta=48, N_\phi=32$. The
dotted-dashed lines show examples of minimizations at lower
grid resolution and $N_\theta=64, N_\phi=48$.}
\label{fig:MOTS}
\end{figure}

Figure~\ref{fig:AHconv} shows the residual of the minimization for
various values of $L$ and different separations $\beta$.  The solid
lines represent configurations at different separations of the
holes. They are labeled by the background separation of the holes,
$\beta$.  Each solid line represents a set of minimizations with
varying expansion order $2\le L\le 16$ on the {\em same} initial data
set. At large separations, $\beta\ge 4.5$, the residual of the
minimization becomes independent of $L$ for large $L$. At small
separation, $\beta=4.4$, the residual decreases exponentially through
all computed expansion orders up to $L=16$ -- a common AH has
formed.

Neither reducing the number of rays, nor decreasing the resolution of
the Hamiltonian solver changes the convergence behavior
significantly. This is illustrated by some examples in
Fig.~\ref{fig:AHconv}. We conclude that for this particular example a
common AH first forms between $\beta=4.4$ and $\beta=4.5$.

Expansions to high order in $L$ are essential for discerning the
formation of a common AH.  If one had Fig.~\ref{fig:AHconv}
only up to expansions up to $L=8$, it would be impossible to decide
where the common AH first forms. One would probably conclude
that the common AH forms at larger separations than it actually
does.


\end{multicols}

\newpage
\mediumtext
\begin{table}
\caption{\label{tab:ISCO}
Orbital parameters of the innermost stable
circular orbit for equal-mass spinning holes. The second through sixth
columns give the data obtained in this work, the three columns to the
right give the data for a test mass orbiting a Kerr black hole.  The
results for the $\pps0.08$ and $\pps0.17$ sequences
will have larger systematic errors than the other cases (see text).}
\begin{tabular}{rddddd|ddd}
sequence & $\ell/m$ & $E_b/\mu$  &  $m\Omega$  &  $J/\mu m$
& $L/\mu m$
& $E_b/\mu$ & $L/\mu m$ & $m\Omega$
\\
\tableline
\mms0.50 & 7.05 & -0.0628 & 0.100 & 2.438 & 3.438 & -0.04514 & 3.8842 & 0.04935\\
\mms0.37 & 6.68 & -0.0687 & 0.107 & 2.595 & 3.335 & -0.04767 & 3.7834 & 0.05319\\
\mms0.25 & 6.17 & -0.0743 & 0.120 & 2.730 & 3.230 & -0.05032 & 3.6856 & 0.05727\\
\mms0.12 & 5.58 & -0.0815 & 0.139 & 2.865 & 3.105 & -0.05363 & 3.5738 & 0.06242\\
\pps0.0   & 4.94 & -0.0901 & 0.166 & 2.976 & 2.976 & -0.05719 & 3.4641 & 0.06804\\
\pps0.08  & 4.59 & -0.0975 & 0.186 & 3.042 & 2.882 & -0.05991 & 3.3870 & 0.07237\\
\pps0.17  & 3.93 & -0.1087 & 0.235 & 3.103 & 2.763 & -0.06337 & 3.2957 & 0.07793\\
\end{tabular}

\end{table}

\narrowtext

\begin{table}\caption{\label{tab:AH}
Summary of the common apparent horizon searches.  Listed are the
sequences and values of orbital angular momentum for which an apparent
horizon search was carried out.  The apparent horizon was found to form at a
separation $\ell_1/m<\ell/m<\ell_2/m$.}
\begin{tabular}{lddd}
Sequence & $L/\mu m$ & $\ell_1/m$ & $\ell_2/m$\\
\mms0.37 & 3.38 & 2.32 & 2.38 \\
\mms0.25 & 3.10 & 2.20 & 2.25 \\
\mms0.25 & 3.34 & 2.24 & 2.29 \\
\pps0.0\tablenote{From \cite{Cook:1992:HSI}, which found a critical separation
$\beta=4.17$. This corresponds to a proper separation of
$\ell/m\approx 1.89$.  } 
& 0.0 & 1.89 &\\
\pps0.0  & 2.94 & 2.08 & 2.13 \\
\pps0.0  & 3.00 & 2.08 & 2.13 \\
\pps0.08 & 2.84 & 2.03 & 2.08 \\
\pps0.08 & 2.92 & 2.03 & 2.08 \\
\pps0.17 & 2.79 & 1.98 & 2.03 \\
\pps0.25 & 2.70 & 1.96 & 2.01 \\
\pms0.25 & 3.00 &      & 
\end{tabular}
\end{table}

\twocolumn

\vspace{-0.2in}

\vspace{-0.2in}

\end{document}